\documentclass[a4paper,12pt]{article}
 
\usepackage{tikz}
\usepackage{soul}
\usetikzlibrary{shapes,positioning}
\usepackage{graphicx}
\usepackage{endnotes}
\usepackage{authblk}                                           % For author affiliations, etc...
\usepackage{amssymb}
\usepackage{hyperref}
\usepackage{color}
\definecolor{darkblue}{rgb}{0.0,0.0,0.3}
\hypersetup{colorlinks,breaklinks,linkcolor=darkblue,urlcolor=darkblue,anchorcolor=darkblue,citecolor=darkblue}

\graphicspath{{../../Documents/Work/Graphics/}}

\let\epsilon\varepsilon
\usepackage{chicago}
\usepackage{bbm}                                               % Fancy blackborld bold
\usepackage{tensor}                                            % For fancy indices.
\usepackage{amsmath}

\usepackage{setspace}                           % For changing the spacing for edits etc... Must come BEFORE call of `hyperref' package.
\usepackage{booktabs} % For nice looking tables.
 \usepackage{multirow}
\usepackage{epigraph}
\usepackage{ulem} 
\usepackage{hyperref}
\usepackage[utf8x]{inputenc}

\usepackage[left=1.2in,right=1.2in,top=1.2in,bottom=1.2in]{geometry}

\title{{\bf What Can We Learn From Analogue Experiments?}}

\author{\bf Karim Th\'ebault\thanks{email: \href{mailto:karim.thebault@gmail.com}{karim.thebault@bristol.ac.uk }}}

\affil{{ Department of Philosophy}\\ University of Bristol}   
\date{\today}

\begin{document}

\maketitle

\begin{abstract}

In 1981 Unruh proposed that fluid mechanical experiments could be used to probe key aspects of the quantum phenomenology of black holes. In particular, he claimed that an analogue to Hawking radiation could be created within a fluid mechanical `dumb hole', with the event horizon replaced by a sonic horizon. Since then an entire sub-field of `analogue gravity' has been created. In 2016 Steinhauer reported the experimental observation of quantum Hawking radiation and its entanglement in a Bose-Einstein condensate analogue black hole. What can we learn from such analogue experiments? In particular, in what sense can they provide evidence of novel phenomena such as black hole Hawking radiation?
\end{abstract}

\newpage
\tableofcontents

\section{Introduction}
\subsection{Types of Evidence}

Different modes of inference can be understood as producing different \textit{types of evidence} in various senses.\footnote{In our view, there is not presently available any fully philosophically satisfactory account of evidence in science. Similar sentiments are expressed by \shortcite{cartwright:2010}. Here we offer our own proto-account based upon intuitions regarding the different types of evidence. More complete accounts, inadequate to our present purposes, are \cite{achinstein:2001} and \cite{roush:2005}.} One sense is pragmatic: evidence produced by different modes of inference can be split into different \textit{practical types} in terms of the different forms of action it rationally licences. Evidence based upon highly speculative inferences is of a different practical type to evidence based upon highly reliable inferences in that, for example, it cannot rationally licence risky actions. Even if one is a professional astrologer, one should not decide to cross a busy road whilst blindfolded based upon ones horoscope! As in everyday life, so in science. Evidence based upon principles of parsimony or judgments of mathematical beauty is of a different practical type to that based upon inductive generalisation of appropriately validated experimental results. Whilst both have significant heuristic significance for science, only the latter can be defended as a rational basis for the design of a nuclear power plant. This much is difficult to contest. 

Problems begin when we consider modes of inference that are neither highly speculative nor provenly reliable. Take computer simulation. In what sense do computer simulations produce a different type of evidence to experiments? Are simulations any less a reliable guide to action than experiments? In some cases it seems not. We can and do build things based upon the evidence of computer simulations. We might not, however, think about computer simulations as producing evidence of a form that can  \textit{confirm} a scientific theory. Rather such an evidential role is often taken to be the providence of experiment. That is, although, in terms of the practical actions it licences, evidence based upon computer simulations might be taken to be of the same type as that based upon experiments, prima facie, the two types of evidence should be understood as being of different \textit{epistemic types}. Prima facie, only evidence drawn from experiment can be taken to be of the type that is potentially confirmatory.\footnote{There are, in fact, good arguments that evidence from computer simulation and experiment \textit{are} of the same epistemic type. See \cite{parker:2009,winsberg:2009,beisbart:2012} for discussion of this issue in the literature.} This creates a particular problem when we are in a situation where experimental evidence is hard to come by. How do we test a theory when we are unable to perform experiments to probe the class of phenomena to which its predictions relate? 

One idea, much discussed in the recent literature, is that we can obtain evidence that is of a (potentially) confirmatory epistemic type by \textit{non-empirical} means. For example, arguments based on the observation of an absence of alternative theories are claimed to, in certain circumstances, confer confirmatory support upon theories, such as String Theory, whose predictions we cannot currently directly empirically test \shortcite{dawid:2013,dawid:2014}. A different approach to the problem confirming theories beyond the reach of direct empirical testing is to consider analogue experiments. In this paper we will consider the idea of using \textit{analogue experiments} as a means of performing an \textit{ersatz empirical test} of an important untested theoretical predictions of modern physics. We will consider whether or not evidence gained from analogue experiments can be of the same epistemic type as evidence gained from conventional experiment. We will do this by comparison with evidence produced by speculative inferences that is of a type that cannot be understood as confirmatory: evidence based upon arguments by analogy. We will largely set aside questions concerning the characterisation of confirmation itself.\footnote{Major approaches to confirmation theory (according to a relatively standard classification) are: confirmation by instances, hypothetico-deductivism, and probabilistic or Bayesian approaches. See \cite{Crupi:2013} for more details. For related discussion of different concepts of evidence see \cite{achinstein:2001}. }  
 
\subsection{Analogue Experiments}

Our story starts with one of the most celebrated theoretical results of twentieth century physics: Hawking's \citeyear{hawking:1975} semi-classical argument that associates a radiative flux to black hole event horizons. Although almost universally believed by contemporary theoretical physicists, testing of this prediction of gravitational Hawking radiation is verging on the physically impossible. The temperature of Hawking radiation associated with a solar mass black hole is of the order of one hundred million times smaller than the temperature of the cosmic microwave background. Trying to detect astrophysical Hawking radiation in the night's sky is thus like trying to see the heat from an ice cube against the background of an exploding nuclear bomb. Short of the construction of terrestrial micro-black holes \cite{Scardigli:1999,Dvali:2008}, any \textit{direct} experimental test of Hawking's prediction is effectively physically impossible.

The genesis of analogue gravity comes from the problem of trying to \textit{indirectly} test Hawking's argument. Inspired by the analogy with sound waves in a waterfall, \citeN{Unruh:1981} showed that Hawking's semi-classical arguments can be applied to sonic horizons in fluids. This work spurred the creation of an entire sub-field of `analogue gravity' that features theoretical, and in some cases experimental, condensed matter analogues to a variety of systems including both Schwarzschild black holes and the entire universe  \shortcite{barcelo:2005}. The most crucial experimental results thus far have been achieved by Steinhauer \citeyear{Steinhauer:2014,steinhauer:2016} working with Bose-Einstein condensates at the Technion in Israel. The experiment reported in the 2016 paper is particularly significant and, as discussed in Section \S \ref{exp}, can reasonably be interpreted as the first conclusive confirmation of the existence of quantum Hawking radiation in an analogue system. 

It remains to seen, however, what such analogue experiments in fact tell us about gravitational systems. In particular what kind of evidence regarding the `target systems' (i.e. black holes) do we gain based upon experiments on the `source systems' (i.e. Bose-Einstein condensates). Can analogue experiments provide us with evidence of the same epistemic type as conventional experiments? Or should we think of them as speculative inferences, producing evidence of the same epistemic type as arguments by analogy? In this paper we will attempt to answer such questions, partially drawing inspiration from the work of Dardashti et al. \citeyear{Dardashti:2015,Dardashti:2016}. 

Ultimately, we will conclude that there \textit{is} a plausible theoretical basis to `externally validate' analogue black hole experiments such that a BEC analogue black hole can be taken to `stand in' for an astrophysical black hole. This gives us one example where analogue experiments can provide us with evidence of the same confirmatory epistemic type as conventional experiments. It does not, however, speak to the question of the significance of confirmation. Is it reasonable to think, in quantitive terms, that analogue experiments can provide a comparable degree of confirmation to conventional experiments? Can they be \textit{substantially} confirmatory rather than merely \textit{incrementally} confirmatory? Such questions bring us beyond the scope of the present paper. However, there are reasons to be optimistic. As shown by a recent analysis in terms of Bayesian confirmation theory \shortcite{Dardashti:2016}, given experimental demonstration of an array of analogue Hawking effects across a variety of different mediums the degree of confirmation conferred can be amplified very quickly. It is thus very plausible to think of analogue experiments prospective means for providing confirmatory support that is substantial, rather than merely incremental.
 
 \section{Analogy and Experiment}

Consider the following two examples of successful scientific practice.\footnote{The first example is taken from \cite{Bartha:2013} and the second example is taken from \cite[Appendix 3]{Franklin:2015}.} In 1934 it was observed by a pharmacologist named Schaumann that the compound meperidine had the effect of inducing an S-shaped curved tail when given to mice.  This effect was only previously observed when morphine was given to mice. Together with the similarity in chemical structure, this analogy between the drugs lead Schaumann to reason that meperidine might share morphine's narcotic effects when given to humans. This then proved to be the case. A Bose-Einstein condensate (BEC) is an exotic form of matter that \citeN{Bose:1924} and Einstein \citeyear{einstein:1924,einstein:1925} predicted to exist for a gas of atoms when cooled to a sufficiently low temperature. In 1995, the experimental demonstration of the existence of a BEC was provided using supercooled dilute gases of alkali atoms \shortcite{Anderson:1995}. The crucial observation was a  sharp increase in the density of the gas at a characteristic frequency of the lasers used for cooling. 

The type of inference and type of evidence involved in our two examples are very different. The inference towards Meperidine having narcotic effects is based upon an \textit{argument by analogy}. The inference towards the existence of Bose-Einstein condensation is based upon an \textit{experimental result}. Arguments by analogy are standardly understood as producing evidence of a type that is not capable of providing confirmatory support to scientific claims. Rather they are speculative inferences. Contrastingly, experimental results are standardly understood as producing evidence of a type that is capable of providing confirmatory support to scientific claims. Philosophical analysis sheds light on the reasoning behind such inuitions.  First let us rationally reconstruct the argument by analogy made by Schaumann: 
\begin{itemize}
\item[\textit{P1}.]Morphine is similar to Meperidine on in terms of having similar chemical structure and having the effect of inducing an S-shaped curved tail when given to mice
\item[\textit{P2.}] Morphine has the effect of being a narcotic when given to humans.
\item[\textit{C.}]Therefore, Meperidine will also have the feature of being a narcotic when given to humans
\end{itemize} 
Clearly such an argument is deductively invalid. Moreover, as an inference pattern, it does not met the epistemic standard usually expected of a reliable inductive inference. For this reason we should not think of arguments by analogy as producing evidence of the same epistemic type as reliable inductive inferences. Rather, it seems reasonable to take arguments by analogy to establish only the \textit{plausibility} of a conclusion, and with it grounds for further investigation. This is in line with \citeN{salmon:1990} and Bartha \citeyear{bartha:2010,Bartha:2013}. From this perspective, the importance of analogical arguments is their heuristic role in scientific practice -- they  provide `cognitive strategies for creative discovery' \cite{bailer:2009}. 

Despite their significance and ubiquity within scientific practice the philosophical consensus is that \textit{epistemically speaking} the role of evidence produced by analogical arguments is null.  In contrast, experimental results are usually taken as the key exemplar of epistemically valuable evidence in science. In particular, experimental evidence is the form of evidence that is invoked in the \textit{confirmation} of theories. However, philosophical analysis reveals good reasons to think that experimental results \textit{taken on their own} do not in fact provide epistemic evidence of such unalloyed quality. 
 
Following the work of authors such as \citeN{franklin:1989}, it is valuable to re-consider the epistemological foundations of experimental science.\footnote{See \cite{Franklin:2015} for a full review.} In particular, we can ask questions such as: i) How do we come to believe in an experimental result obtained with a complex experimental apparatus?; and ii) How do we distinguish between a valid result and an artefact created by that apparatus? One of the key ideas in the epistemology of experiment is that to assess the evidence gained from experimentation, we must examine and evaluate the strategies that scientists use to \textit{validate} observations within good experimental procedures. Following \citeN{winsberg:2010} we can make an important distinction between two different types of validation in the context of experimental science: An experimental result is \textit{internally valid} when the experimenter is genuinely learning about the actual system they are manipulating -- when, that is, the system is not being unduly disturbed by outside interferences; An experimental result is \textit{externally valid} when the information learned about the system being manipulated is relevantly probative about the class of systems that are of interest to the experimenters. 

Consider the case of the 1995 experiments that were taken to provide conclusive evidence for the existence of BECs on the basis of experiments on supercooled dilute gases of alkali atoms.  The internal validity of the experiments relates to the question of whether or not the results obtained genuinely reflect the fact that the particular supercooled dilute gases of alkali atoms experimented upon were behaving as BECs. The external validity  of the experiments relates to the question of whether or not the inferences regarding the particular \textit{source systems} experimented upon (particular supercooled dilute gases of alkali atoms) can be reliably generalised to the wide class of \textit{target systems} that the theory of BECs refers to. Experimental results can only be reasonably be taken to constitute evidence that is of an epistemic type that makes it suitable for confirmation on the assumption that they have been both internally and externally validated. While the idea that an experimental result must be internally validated is rather a familiar one, the notion of external validation is not as frequently discussed -- particularly by actual experimental scientists. In the end, external validation is the crucial link from an experimental result to the use of this result as a token of the type of epistemic evidence relevant to the confirmation of general scientific statements. Clearly, an experiment that is not internally validated does not produce evidence of an epistemic type that renders it suitable for confirming any scientific hypothesis. On the other hand, an experiment that is internally validated but not externally validated produces evidence of an  epistemic type that can confirm specific statements regarding the particular system experimented upon, but cannot confirm statements regarding the wider class of systems that the experiment is designed to probe.  As such, theoretical arguments for external validation are almost always going to be a necessary requirement for integrating experimental results into scientific knowledge.  

This brief discussion of analogy and experiment is intended to provide a prospectus for a philosophical analysis of analogue experiments. The key question in the \textit{epistemology of analogue experimentation} is whether there are arguments that can provide a suitable form of external validation. If there are no such arguments, then we \textit{should not} consider analogue experiments as providing evidence regarding the relevant `target systems' (black holes in the Hawking radiation case) that is of the same epistemic kind as (externally validated) experiments. In such circumstances analogue experiments would have a more heuristic role, like arguments by analogy, and could not in principle be used to produce evidence used for the confirmation of general scientific statements regarding the target system. However, if arguments for external validation can be found and justified in cases of analogue experimentation then we \textit{can} legitimately think about analogue experiments as providing evidence of the same epistemic type as conventional (externally validated) experiments. 

Significantly, this question is distinct to any question of the \textit{strength of evidence}. It may be the case, in either a conventional experiment or analogue experiment, that the validation procedure is not highly reliable, in which case the evidence will be of the same epistemic type, but of a different strength. While questions regarding strength of evidence suggests immediately an analysis in terms of Bayesian confirmation theory along the lines of \shortcite{Dardashti:2016}, we hold that questions of type of evidence can be answered independently of such probabilistic modes of analysis.  

\section{The Hawking Effect}

In this section we will present the formal details behind the Hawking effect as instantiated in gravitational systems, fluid dynamical systems and Bose-Einstein condensates. For the most part we will follow the account of \shortcite{barcelo:2005}.   which is also a good source to find supplementary details.

\subsection{Gravitational Hawking Effect}

In a semi-classical approach to gravity we consider a quantum field within a fixed spacetime background. For this modelling framework to be valid it is assumed that we are considering quanta of wavelengths much larger than the Planck length. In the simplest semi-classical model we consider a massless scalar field operator $\hat{\phi}$ that obeys a wave equation of the form $g^{ab}\nabla_a \nabla_b \hat{\phi} = 0$. We can expand the scalar field in a basis of orthonormal plane wave solutions: 
\begin{equation}
\hat{\phi} = \int d\omega   (\hat{a}_\omega f_\omega + \hat{a}^\dagger_\omega f^*_\omega),
\end{equation}
where $ f_\omega = \frac{1}{\sqrt{2}}e^{-i(\omega t- kx)}$ and $\hat{a}_\omega$, $\hat{a}^\dagger_\omega$ are creation and annihilation operators for modes of some frequency $\omega$. The creation and annihilation operators allows us to define both a vacuum state, $\hat{a}_\omega |0\rangle=0$, and a number operator, $\hat{N}_\omega =  \hat{a}_\omega^\dagger \hat{a}_\omega$, in this particular basis. We call a vacuum state defined for the scalar field at past null infinity, $\mathcal{J}^{-}$, the `in' state, and a vacuum state defined for future null infinity, $\mathcal{J}^{+}$, the `out' state. In general, the `in' state need not appear as a vacuum state to observers at positive null infinity: it may contain a flux of `out-particles' which one can calculate simply by determining the Bogoliubov coefficients between the solutions expressed in the `in' and `out' basis:

 \begin{equation} \label{number}
_{in}\langle 0| (\hat{N}^{out}_\omega)  |0\rangle_{in}   
 =  \int  d\omega' |\beta_{ \omega \omega'}|^2.
\end{equation}

What Hawking's 1975 calculation shows is that, for a spacetime which features the establishment of an event horizon via gravitational collapse leading to a black hole, one can derive the asymptotic form of the Bogoliubov coefficients and show that it depends only upon the \textit{surface gravity} of the black hole denoted by $\kappa_{G}$. Surface gravity is the magnitude of the acceleration with respect to Killing time of a stationary zero angular momentum particle just outside the horizon. Hawking's calculation implies that a black hole horizon\footnote{\citeN{giddings:2016} has argued that we should trace the origin of Hawking radiation to a `quantum atmosphere' some distance away from the horizon. The implications of this idea for analogue black hole experiments is an interesting issue to consider.} has intrinsic properties that are connected to a non-zero particle flux at late times. The spectrum of this flux obeys the relation:
\begin{equation}
\langle \hat{N}_{\omega}^{\text{Black Hole}} \rangle = \frac{1}{e^{\frac{2 \pi \omega}{\hbar \kappa_{G}}}-1}\qquad  T_{BH}=\hbar \kappa_{G}/ 2\pi
\end{equation}
Crucially, the functional form of this spectrum is \textit{thermal} in the sense that it takes a characteristic Planckian blackbody energy form for the temperature $T_{BH}$. Black holes, it turns out, are hot!

\subsection{Hydrodynamic Hawking Effect}

Consider a classical fluid as a continuous, compressible, inviscid medium and sound as an alternate compression and rarefaction at each \textit{point} in the fluid. The points are \textit{volume elements} and are taken to be \textit{very small} with respect to the overall fluid volume, and \textit{very large} with respect to the inter-molecular distances. The modelling framework of continuum hydrodynamics is thus only valid provided fluid density fluctuations of the order of molecular lengths can be ignored. The two fundamental equations of continuum hydrodynamics are the \textit{continuity equation}, which expresses  the conservation of matter:
\begin{equation}
\frac{\partial \rho}{\partial t}  +  \nabla \cdot (\rho \vec{v}) = 0,
\end{equation}
and the \textit{Euler equation}, which is essentially Newtons second law: 
 \begin{equation}
\rho \Big( \frac{\partial \vec{v}}{\partial t}  +   (\vec{v}\cdot \nabla)\vec{v}  \Big) = -\nabla p,
\end{equation}
where $\rho$ is the mass density of the fluid at a particular point, $\vec{v}$ is the velocity of the fluid volume element, and $p$ is pressure.

If the fluid is \textit{barotropic} and \textit{locally irrotational} Euler's equation reduces to a form of the Bernoulli equation. We identify sound waves in the fluids with the fluctuations $(\rho_1, p_1, \psi_1)$ about the background, which is interpreted as bulk fluid motion. The linearised version of the continuity equation then allows us to write the equation of motion for the fluctuations as:
\begin{equation} 
\frac{1}{\sqrt{-g}}\frac{\partial}{\partial x_\mu}    ( \sqrt{-g}  g^{\mu \nu }    \frac{\partial}{\partial x^\nu}  \psi_1 )  =0, 
\end{equation}
where we have defined the acoustic metric
\begin{equation}
g_{\mu \nu}^{\text{acoustic}} = \frac{\rho_0}{c_{\text{sound}}}  \left( \begin{array}{ccc}
-(c_{sound}^2-v_0^2) & \vdots & -(v_0)_j    \\ 
\ldots & \cdot & \ldots \\ 
-(v_0)^i  & \vdots & \delta_{ij}
\end{array} \right).
\end{equation}

Propagation of sound in a fluid can be understood as being governed by an acoustic metric of the form $g_{\mu \nu}$. The close similarity between the acoustic case and gravity can be seen immediately if we consider the Schwarzschild metric in Painleve-Gullstrand coordinates:

\begin{equation}
g_{\mu \nu}^{\text{Schwarzschild}} =   \left( \begin{array}{ccc}
-(c_0^2-\frac{2 GM}{r}) & \vdots & -\sqrt{\frac{2 GM}{r}}\vec{r_j}    \\ 
\ldots & \cdot & \ldots \\ 
-\sqrt{\frac{2 GM}{r}}\vec{r_i}  & \vdots & \delta_{ij}
\end{array} \right).
\end{equation}
 This similarity can be transformed into an isomorphism (up to a factor) given certain conditions on the speed of sound in the fluid and the fluid density and velocity profiles. The role of the black hole event horizon is now played by the effective acoustic horizon where the inward flowing magnitude of the radial velocity of the fluid exceeds the speed of sound.  The black hole is replaced by a \textit{dumb hole}.

Unruh's crucial insight in his 1981 paper was that once the relevant fluid-spacetime geometric identification has been made, there is nothing stopping one from repeating Hawking's 1975 semi-classical argument, only replacing light with sound. The result is that while, in the gravitational Hawking effect a black hole event horizon is associated with a late time \textit{thermal photonic flux}, in the hydrodynamic Hawking effect a dumb hole sonic horizon can be associated with a late time \textit{thermal phononic flux}.

\subsection{BEC Hawking Effect}

Following the same line of reasoning as Unruh's original ideal fluid argument, \shortciteN{garay:2000} derived a  BEC Hawking Effect using appeal to the hydrodynamic approximation of a BEC. Consider the Gross-Pitaevskii equation that can be derived by applying a mean field approximation to the many body QM description of a BEC:
\begin{eqnarray}\label{eq:GP}
i\hbar \frac{\partial \psi(\mathbf{r},t)}{\partial t}&=&-\frac{\hbar^2}{2m}\nabla^2\psi(\mathbf{r},t)+V(\mathbf{r})\psi(\mathbf{r},t)+U_0 |\psi(\mathbf{r},t)|^2\psi(\mathbf{r},t) \\ &=&H_{GP}\psi(\mathbf{r},t),
\end{eqnarray}
where $V(\mathbf{r})$ is an external potential. $U_0=4\pi\hbar^2a/m$ is the effective two two particle interaction, with $a$ and $m$ the scattering length and atomic mass.
From this one can obtain a Madelung-type expression of the form: 
\begin{equation}\label{eq:becEE}
\frac{\partial \mathbf{v}}{\partial t} = -\frac{1}{nm} \nabla p - \nabla \Big(\frac{v^{2}}{2}\Big)+\frac{1}{m} \nabla \Big(\frac{\hbar^{2}}{2m\sqrt n}\nabla^{2}\sqrt n \Big)-\frac{1}{m}\nabla V,
\end{equation}
where $\psi=fe^{i\phi}$, $n=|\psi|^2$, $p=n^{2}U_{0}/2$,  and $\mathbf{v}=\frac{\hbar}{m}\nabla \phi$. Consider the quantum pressure term:
\begin{equation*}
\frac{1}{m} \nabla \Big(\frac{\hbar^{2}}{2m\sqrt n}\nabla^{2}\sqrt n \Big).
\end{equation*}
When variations in the density of the BEC occur on length scales much greater than the \textit{healing length}, $\xi=\hbar/(mnU_{0})^{\frac{1}{2}}$, the quantum pressure term can be neglected and we recover the usual classical Euler equation for an irrotational fluid. Now consider a `second quantised' field theoretic description of the (weakly interacting) BEC, where the Hamiltonian is expressed in terms of creation and annihilation operators for Bosons, $\hat{\psi}^\dagger(\mathbf{r})$ and $\hat{\psi}(\mathbf{r})$. We can decompose the quantum field into a classical `bulk' field and a quantum fluctuation:
\begin{equation}
\hat{\psi}(\mathbf{r},t)= \psi(\mathbf{r},t) +\hat{\phi}(\mathbf{r},t).
\end{equation}
The Gross-Pitaevskii equation can be recovered as the equation for the classical field $\psi(\mathbf{r},t)$, when the backreaction with the quantum fluctuation can be neglected. What \shortciteN{garay:2000} showed was that in the limit where we have no backreaction and the quantum pressure term  can be neglected, linearized fluctuations in the BEC will couple to an effective acoustic metric of the same form as that derived by Unruh for the hydrodynamic case. In this limit the derivation of the BEC Hawking effect then follows the same pattern as that for hydrodynamics and gravity. 

In the Bogoliubov description of a (weakly interacting) BEC one assumes that the backreaction of the classical field with the quantum excitations is  \textit{small but non-negligible} and includes only terms which are at most quadratic in $\hat{\phi}(\mathbf{r},t)$.\footnote{See \citeN{pethick:2002} for a detailed textbook treatment of the Bogoliubov description of BECs.} In the Bogoliubov description fluctuations are no longer governed by an effective acoustic metric, rather elementary excitations correspond to eigenvalues of the Bogoliubov operator: 
\begin{equation}
\mathcal{L} =   \left( \begin{array}{cc}
H_{GP} - \mu+  U_0 |\psi|^2 &  U_0 \psi^2      \\ 
- U_0 \psi^{\star2} & -H_{GP} + \mu-  U_0 |\psi|^2 
\end{array} \right),
\end{equation}
where $\mu$ is the chemical potential. Impressively, it has been shown by \shortciteN{recati:2009} that one can still derive an analogue Hawing effect for an inhomogeneous BEC when treated in the Bogoliubov description. This particularly interesting since in this regime there is no longer technically an analogy with the gravitational case since the `surface gravity' is formally infinite. The authors note:   
\begin{quote}
It is remarkable to note that [the results] still give the typical thermal behaviour of Hawking radiation even though one is not allowed to use the gravitational analogy. \shortcite[p.6]{recati:2009}
\end{quote}
There is thus an interesting sense in which the BEC Hawking effect is \textit{robust} to the breakdown in the hydrodynamic description at lengths scales smaller than the healing length. 

\section{The Trans-Planckian Problem and Universality}

In the standard calculation of the Hawking temperature exponential gravitational red-shift means that the black hole radiation detected at late times (i.e. the outgoing particles) must be taken to correspond to extremely high frequency radiation at the horizon.  Such a `trans-Planckian' regime is the dominion of theories of quantum gravity, and is thus well beyond the domain of applicability of the modelling framework we are using. This problem with `trans-Planckian' modes has a direct analogue in the BEC case in terms of the failure of the hydrodynamic limit -- we cannot assume that the perturbations have wavelengths much larger than the healing length. Of course since the results of \shortciteN{recati:2009} show that the BEC Hawking effect \textit{is} suitably robust, we have good reasons not to worry too much about the `trans-Planckian' problem for the BEC analogue model. But whereas the Bogoliubov description gives us microphysical understanding of a BEC, we do not have an equivalently trustworthy theory for the microphysics of spacetime or, for that matter, fluids. 

We can, however, attempt to model the effect of the underlying microphysics on linear fluctuations by considering a modified dispersion relation. This idea originally comes from Jacobson \citeyear{Jacobson:1991,Jacobson:1993}, who suggested that one could use a modified dispersion relation to understand the breakdown of continuous fluid models due to atomic effects. The question of particularly importance is whether or not an exponential relationship actually holds between the outgoing wave at some time after the formation of the horizon, and the wavenumber of the wave packet \cite{unruh:2008}. Using numerical simulations it was first shown by \citeN{unruh:1995} that the altered dispersion relation in atomic fluids does imply that the early time quantum fluctuations that cause the late-time radiation are not in fact exponentially large. A related analytical argument was later applied to the gravitational case by \citeN{corley:1998}. What is more desirable, however, is a set of general conditions under which such effective decoupling between the sub- and trans- Planckian physics can be argued to take place. One interesting proposal in this vein is due to \citeN{unruh:2005}.\footnote{For further work on these issues, using a range of different methodologies, see \shortciteN{himemoto:2000}, \shortciteN{barcelo:2009} and \shortciteN{coutant:2012}.} The `universality' results of Unruh and Sch\"{u}tzhold  show that possible trans-Planckian effects can be factored into the calculation of Hawking radiation via a non-trivial dispersion relation. In particular, they consider a generalised Klein-Fock-Gordon equation for the scalar field of the form:
\begin{equation}
\bigg(\partial_t +\partial_{x}v(x)\bigg)\bigg(\partial_t +v(x)\partial_{x}\bigg)\hat{\phi}= \bigg( \partial^{2}_x+F(\partial^{2}_x)\bigg) \hat{\phi}
\end{equation}
with $F$  the non-trivial dispersion relation and $v$ \textit{either} the local velocity of free falling frames measured with respect to a time defined by the Killing vector of the stationary metric \textit{or} the position dependent velocity of the fluid. They use complex analytic arguments to show that the Hawking flux is robust under the modified dispersion relation -- essentially this is because the dominant terms in the relevant integral are entirely due to the discontinuity caused by the horizon of the black hole (or its analogue). What Unruh and Sch\"{u}tzhold establish is that the Hawking effect does not, to lowest order, depend on the details of underlying physics, given certain modelling assumptions.\footnote{For example, the evolution of the modes is assumed to be adiabatic -- the Planckian dynamics is supposed to be much faster than all external (sub-Planckian) variations} 

\section{The Technion Experiments}\label{exp}

Based upon the most recent of a series of experiments\footnote{Three particular landmark experimental results achieved by Steinhauer (and in the first case collaborators) are the creation of a BEC analogue black hole \shortcite{lahav:2010}, the observation of self-amplifying Hawking radiation in an analogue black-hole laser \cite{Steinhauer:2014}  and the observation of quantum Hawking radiation and its entanglement \cite{steinhauer:2016}. } at the Technion (Israel Institute of Technology),  Steinhauer has claimed to have observed BEC Hawking radiation. This observation is based upon measurement of the density correlations in the BEC that are understood to `bear witness' to entanglement between Hawking modes outside the sonic horizon and partner modes inside the sonic horizon. In this section we will provide some background to these results before concluding our analysis by considering the potential for the Technion experiments to function as evidence for gravitational Hawking radiation. 

The first key idea we must introduce in order to understand the theoretical context of Steinhauer's work is the idea of entanglement as a `witness' to Hawking radiation.  It is plausible to think of the Hawking effect for black holes in the terms of creation of correlated pairs of outgoing quanta triggered by the event horizon formation \shortcite{balbinot:2008,schutzhold:2010,parentani:2010}.  The particles that featured in our treatment of the Hawking effect in \S3 correspond to escaping modes that make up one half of each pair of quanta. These `Hawking modes' are each correlated with (rather ill-fated) `partner modes'  that are trapped inside the event horizon and forever lost. It is thus understandable that in the gravitational context the Hawking modes are the centre of attention. Contrastingly, when considering the Hawking effect in an analogue context the regions separated by the horizon are not causally disconnected: unlike in the case of an event horizon, in principle an observer can have access to both sides of a sonic horizon. The key idea is that by measuring correlations between modes inside and outside horizon, we can detect the signature of Hawking radiation in an analogue black hole. In particular, if we can establish that there is entanglement between the modes, then we have strong evidence for the existence for \textit{quantum} Hawking radiation within the relevant analogue system.   

The next important idea is that in the case of a BEC analogue black hole the entanglement between the Hawking and partner modes can be measured by considering the density-density correlation function between spatial points either side of the horizon \cite{steinhauer:2015}. This is highly important from a practical perspective since there exist high precision techniques for measuring point by point densities of a BEC by considering phase shifts of a laser beam shone through it. Formally, the relationship between the entanglement of the modes and the density correlations can be established by re-writing a given measure of the non-separability of the quantum state $\triangle$, in terms of density operators $\rho_k$, for modes of wavenumber $k$.  For example, in his treatment Steinhauer \citeyear{steinhauer:2015,steinhauer:2016} considers the simple measure:
\begin{equation}
\triangle \equiv \big{<} \hat{b}^{u\dag}_{k_{H}} \hat{b}^{u}_{k_{H}}\big{>}\big{<} \hat{b}^{d\dag}_{k_{p}} \hat{b}^{d}_{k_{p}}\big{>}-\big{|} \hat{b}^{u}_{k_{H}} \hat{b}^{d}_{k_{p}}\big{|}, 
\end{equation} 
where $\hat{b}^{u}_{k_{H}}$ is the annihilation operator for a Bogoliubov excitation with wave number $k_{H}$, localised in the subsonic `downstream' region outside the event horizon. The letters $u$, $d$, $H$ and $p$ thus stand for upstream (supersonic), downstream (subsonic), Hawking modes and partner modes respectively. Using the Fourier transform of the density operator we  can re-express such a non-separability measure in terms of correlations between $u$ and $d$ density operators for the $k_{H}$ and $k_p$ modes. We can thus measure $\triangle$ purely by measuring density correlations in the BEC on either side of the sonic horizon. If $\triangle$ is negative then the correlations are strong enough to indicate entanglement, and thus the quantum signature of Hawking radiation. 

In his landmark experiment Steinhauer used a BEC of $^{87}$Rb atoms confined radially by a narrow laser beam. The horizon was created by a very sharp potential step which is swept along the the BEC at a constant speed. Significantly the length scales are such that the hydrodynamic description of a BEC is appropriate: the width of the horizon is of the order of a few times bigger than the healing length. The main  experimental result consists of an aggregate correlation function computed based upon an ensemble of 4,600 repeated experiments which were conducted over six days. Given some reasonable assumptions (for example modes at different frequency are assumed to be independent of each other) the experiments can be interpreted as establishing an  entanglement witness to Hawking radiation in  BEC. 

\section{Dumb Hole Epistemology}

Three notions of validation are relevant to the Technion experiments described above. The first, and most straightforward, is internal validation. Was Steinhauer genuinely learning about the physics of the particular sonic horizon within the particular $^{87}$Rb BEC that he was manipulating? Various sources of internal validation are apparent from the description of the experimental set up given, not least the repetition of the experimental procedure nearly five thousand times. Given this, the evidence gained from from the experiments conducted can be categorised as of the appropriate epistemic type to be used to confirm specific statements regarding the particular BEC that was experimented upon. The next question relates to external validation of the Technion experiments as experiments in the conventional sense. Can the particular sonic horizon that was constructed, within the particular $^{87}$Rb BEC, stand in for a wider class of systems -- for example, all BEC sonic horizons within the realm of validity of the hydrodynamic approximation to the Gross-Pitaevskii equation, regardless of whether the relevant systems have been (or even could be) constructed on earth. Given this set of systems obeys the `reasonable assumptions' of the Steinhauer experiments, such as modes at different frequency are assumed to be independent of each other, then we can also externally validate the experiments. This means, with relevant qualifications, the evidence produced by the experiments conducted can be categorised as of the appropriate epistemic type to be used to confirm general statements regarding the class of BECs with sonic horizons. Thus, so far as the status as conventional experiments go, it reasonable to take the Technion experiments to be internally and externally validated based upon purely upon Steinhauer's report. 

The details provided in the report do not, however, function as external validation of the Technion experiments \textit{as analogue experiments}. That is, without further argument we do not have a link from the class of source systems (BECs with sonic horizons) to the class of target systems (astrophysical black holes). Thus, when considered in isolation the evidence from the Technion experiments cannot be categorised as of the appropriate epistemic type to be used to confirm general statements regarding astrophysical black holes. Rather, when considered in isolation, regarding such statements, it is of the same epistemic type as evidence derived from speculative inferences. Although it might reasonably be argued to be evidence that is, in some sense, more convincing or valuable than that produced by many such inferences, in confirmatory terms it is equally null.  

The question of external validation of dumb hole experiments as as analogue experiments is where the universality arguments of Unruh and Sch\"{u}tzhold can be brought in. If accepted, the theoretical universality arguments of Unruh and Sch\"{u}tzhold would function as external validation for the Steinhauer experiments. They provide a theoretical basis for taking the source system of the Technion experiments (a BEC analogue black hole) to stand in for a wider class of target systems (black holes in general including other analogue models and gravitational black holes). In this sense, they give us a possible basis to upgrade the epistemic type of the experimental evidence, such that it can be used to confirm general statements regarding astrophysical black holes.  

This should not, perhaps, be as surprising a statement as it sounds. After all, almost by definition, this is what universality arguments mean: they are theoretical statements about the multiple-realisability of a given phenomenon, and so imply that certain features of the phenomenon in one exemplification will be present in all others. That said, it is difficult not to be wary of the speed and strength of this kind of conclusion. This mode of external validation is almost completely unlike those used in more conventional experiments and, as such, it should be treated rather sceptically for the time being.  Arguably, the importance of the Technion experiments lies in their future rather than immediate evidential value. In particular, given experimental demonstration of an array of analogue Hawking effects across a variety of different mediums, one could become increasingly confident in the universality arguments and thus, in turn, in the external validation of the experiments as analogue experiments.\footnote{There are grounds for optimism in this regard on a number for fronts. See \shortciteN{philbin:2008,belgiorno:2010,unruh:2012,liberati:2012,nguyen:2015}.} 

We can thus conclude that there \textit{is} a plausible theoretical basis to `externally validate' analogue black hole experiments such that a BEC analogue black hole can be taken to `stand in' for an astrophysical black hole. This gives us one example where analogue experiments can provide us with evidence of the same confirmatory epistemic type as conventional experiments. It does not, however, speak to the question of the significance of confirmation. Is it reasonable to think, in quantitive terms, that analogue experiments can provide a comparable degree of confirmation to conventional experiments? Can they be \textit{substantially} confirmatory rather than merely \textit{incrementally} confirmatory? Such questions bring us beyond the scope of the present paper. However, there are reasons to be optimistic. As shown by a recent analysis in terms of Bayesian confirmation theory \shortcite{Dardashti:2016}, given experimental demonstration of an array of analogue Hawking effects across a variety of different mediums the degree of confirmation conferred can be amplified very quickly. It is thus very plausible to think of analogue experiments prospective means for providing confirmatory support that is substantial, rather than merely incremental.    

\section*{Acknowledgement}

I am deeply indebted to Radin Dardashti, Richard Dawid, Stephan Hartmann, and Eric Winsberg for discussions and collaborative engagement without which this paper could not have been written. I am also thankful to the audience at the \textit{Why Trust a Theory?}  workshop in Munich for insightful questions, to Bill Unruh for providing me with some useful details about the Technion experiments, and to Erik Curiel for various helpful discussions about black hole thermodynamics.

\bibliographystyle{chicago}
\bibliography{dumb}

\end{document}